\documentclass[letterpaper,twocolumn,prb,aps,showpacs,superscriptaddress,floatfix]{revtex4-1}
\usepackage[latin1]{inputenc}
\usepackage{bm}
\usepackage[usenames]{color}
\usepackage{multirow}
\usepackage{amssymb}
\usepackage{amsbsy}
\usepackage{amsmath}
\usepackage{stmaryrd}
\usepackage{graphicx}
\usepackage{epsfig}
\usepackage{placeins}
\usepackage{ulem}
\makeatletter

\begin{document}

\title{Energy Dynamics in the Heisenberg-Kitaev Chain}

\author{Robin Steinigeweg}
\email{rsteinig@uos.de}
\affiliation{Department of Physics, University of Osnabr\"uck, D-49069 Osnabr\"uck, Germany}
\affiliation{Institute for Theoretical Physics, Technical University Braunschweig, D-38106 Braunschweig, Germany}

\author{Wolfram Brenig}
\affiliation{Institute for Theoretical Physics, Technical University Braunschweig, D-38106 Braunschweig, Germany}

\date{\today}

\begin{abstract}
We study the Heisenberg-Kitaev chain in order to uncover
the interplay between two qualitatively different integrable points
in the physics of heat transport in one dimension. Focusing on high
temperatures and using analytical as well as numerical approaches within
linear response theory, we explore several directions in parameter
space including exchange-coupling ratios, anisotropies, and external
magnetic fields. We show the emergence of purely ballistic energy
transport at all integrable points, manifest in pronounced Drude
weights and low-frequency suppression of regular-conductivity
contributions. Moreover, off integrability, we find extended quantum
chaotic regions with vanishing Drude weights and well-defined DC
conductivities. In the vicinity of the Kitaev point, we observe clear
signatures of the topological gap in the response function. This gap
coexists with a nonzero Drude weight in the Kitaev chain.
\end{abstract}


\maketitle

\section{Introduction}

Integrable quantum many-body models serve as idealized testbeds to study
thermodynamic and dynamical properties at both, zero and non-zero
temperatures. A prominent example in this context is Kitaev's exact
solution of a quantum spin model on the honeycomb lattice with strong
exchange anisotropy \cite{Kitaev2006}. This model harbors a spin
liquid in two dimensions, with either gapped or gapless emergent
Majorana-fermion excitations, $Z_2$ gauge fluxes, and a field-induced
phase with non-Abelian quasi-particle excitations \cite{Kitaev2006,
Jiang2011}. Early on, Mott-insulating layered iridates A$_2$IrO$_3$,
which display strong spin-orbit coupling, have been suggested as
promising material candidates for Kitaev's model. However, additional
$SU(2)$ invariant Heisenberg exchange interactions beyond the bare
Kitaev model have also been realized as an inevitable ingredient in
any realistic context \cite{Jackeli2009, Chaloupka2010, Singh2012}.

Remarkably, the notion of Majorana particles withstands finite
temperatures and the truncation of Kitaev's model via $n$-leg ladders
down to one-dimensional (1D) chains \cite{Kitaev2001, Feng2007, Hu2012}, where
remnants of gauge-field physics can remain active through an extensive
number of conservation laws \cite{Saket2013}. On the one hand, the
dynamics of bare Kitaev chains has met an upsurge of interest in the
context of transport through 1D topological superconductors
\cite{Kitaev2009, Alicea2012}, where Majorana edge modes may have been
observed \cite{Mourik2012}. On the other hand, transport through pure
Heisenberg chains has come under intensive scrutiny ever since the
observation of colossal spinon heat-conduction in 1D cuprates
\cite{Sologubenko2000, Hess2007, Hlubek2010} and the discovery of
diverging transport coefficients in $S=1/2$ Heisenberg chains
\cite{Zotos1997, Zotos1999, Kluemper2002, FHM2003}.

Integrability of Heisenberg chains has surfaced the key to anomalous
spinon transport. Notably, the spin heat-current is a strictly conserved
quantity of the XXZ chain at any temperature, which results in a
non-zero Drude weight for heat transport \cite{Kluemper2002, FHM2003}
and is in line with the large heat conduction observed experimentally
in almost ideal realizations of the XXZ model \cite{Sologubenko2000,
Hess2007, Hlubek2010}. However, the spectrum of the spin-current
response, including the question of a non-zero Drude weight for spin
transport, remains an issue only partially solved at non-zero
temperatures even after more than two decades \cite{Shastry1990,
Zotos1997, Zotos1999, Rosch2000, FHM2003, Benz2005, Sirker2009,
Grossjohann2010, Prosen2011, Steinigeweg2011, Herbrych2012,
Karrasch2013}.

Due to its singular nature, transport in the XXZ chain is highly
susceptible to integrability-breaking interactions, e.g.,
further-neighbor exchange \cite{FHM2003, FHM2004, Jung2007}, lattice
degrees of freedom \cite{Shimshoni2005, Rozhkov2005, Hlubek2012}, and
disorder \cite{Karahalios2009}. In this context Kitaev's anisotropic
exchange is rather remarkable since it allows to tune between two
completely different integrable points, namely, the Heisenberg and
the Kitaev chain. Yet, the interplay between these two types of
exchange interactions regarding transport properties of spin chains
has never been studied, neither for zero nor for non-zero temperatures.

In this paper, we study energy transport in the Heisenberg-Kitaev
chain, focusing on high temperatures as a first step. Using analytical
and numerical approaches within linear response theory, we explore
several directions in parameter space. As our main finding, we uncover
ballistic transport with reduced low-frequency spectral weight at all
integrable points, embed in a range of extended quantum chaotic regions
with well-defined DC conductivities. In the vicinity of the Kitaev
point, we numerically observe signatures of the topological gap in
the response function. This gap coexists with a nonzero Drude weight
in the Kitaev chain.

The paper is structured as follows: In the next Sec.\ \ref{kitaev}
we study the bare Kitaev chain. We first discuss the Hamiltonian and
the associated energy-current operator in Sec.\ \ref{energycurrent} and
then derive in Sec.\ \ref{autocorrelation} an analytical expression
for the energy-current autocorrelation in the thermodynamic limit. In
Sec.\ \ref{heisenbergkitaev} we extend our study to the Heisenberg-Kitaev
chain and numerically analyze the role of XXZ interactions. To this end,
we perform exact diagonalization for various exchange-coupling ratios and
anisotropies. In Sec.\ \ref{magneticfield} we further analyze the role of
an external magnetic field. We close in Sec.\ \ref{conclusion} with
summary and conclusion.

\section{Kitaev Model}
\label{kitaev}

\subsection{Hamiltonian and Energy Current}
\label{energycurrent}

We begin with the bare Kitaev chain, for which we obtain analytic
results. In one dimension, and with periodic boundary conditions, its
Hamiltonian reads
\cite{Kitaev2001, Feng2007}
\begin{equation}
H = \sum_{l=1}^{L/2} h_l \, , \quad h_l = J_1 \, S_{2l-1}^x S_{2l}^x
+ J_2 \, S_{2l}^y S_{2l+1}^y \, ,
\end{equation}
where $L$ is an even number of sites, $S_r^x$, $S_r^y$ are the $x$,
$y$ components of spin-1/2 operators at site $r$, and $J_1, J_2 \in
\mathbb{R}$ are exchange coupling constants. The unit cell contains
two bonds. We note that Kitaev's chain allows for $L/2$ mutually
commuting $Z_2$ invariants $S_{2l}^x S_{2l+1}^x$ ($S_{2l-1}^y S_{2l}^y$),
$l = 1, 2, \ldots L/2$.

The energy current follows from the continuity equation
\cite{Zotos1997}
\begin{equation}
j = \sum_{l=1}^{L/2} \imath [h_l,h_{l+1}] = J_1 J_2
\sum_{l=1}^{L/2} S_{2l}^y S_{2l+1}^z S_{2l+2}^x \, .
\label{Eq2}
\end{equation}
This operator acts on three adjacent sites and contains also
the $z$ component of spin-$1/2$ operators.

Both, the Hamiltonian and the energy current can be brought
into a spinless-fermion representation using the Jordan-Wigner
transformation. At that point, the Hamiltonian can also be rewritten
as a model of Majorana fermions \cite{Feng2007}, where Eq.~(\ref{Eq2})
refers to their associated energy flow. Fourier and Bogoliubov
transformation diagonalizes the spinless-fermion Hamiltonian into
\begin{equation}
H = \sum_{k=0}^{\pi/2} \, [H_1(k) c_k^\dagger c_k + H_2(k) d_k^\dagger
d_k]
\end{equation}
with a one-particle dispersion
\begin{equation}
H_{i \in \{1,2\}}(k)= \frac{(-1)^i}{2} \sqrt{J_1^2 + J_2^2 + 2 J_1 J_2
\cos 2 k}
\end{equation} consisting of two branches
for $c_k^{(\dagger)}$ and $d_k^{(\dagger)}$ fermions \cite{Feng2007,
Chen2008}. We note that the momentum $k = 2\pi l/L$, $l=1,2,\ldots,
L/4$ is confined to $k \leq \pi/2$ and that two `zero modes'
$H_{i \in \{3,4\}}(k) = 0$ exist.

The heat current in Eq.~(\ref{Eq2}) can also be written in terms
of $c_k^{(\dagger)}$ and $d_k^{(\dagger)}$, but will be non-diagonal
in this representation. E.g.~for $J_1 = J_2$, the Hamiltonian is
\begin{equation}
H = J_1 \sum_{k=0}^{\pi/2} \cos k \, (c_k^\dagger c_k - d_k^\dagger
d_k) \, ,
\end{equation}
however, the energy current reads
\begin{equation}
j = \frac{J_1^2}{4} \, \sum_{k=0}^{\pi/2} \, \sin 2 k \,
(c_k^\dagger c_k + d_k^\dagger d_k + c_k^\dagger d_k + d_k^\dagger
c_k) \, .
\end{equation}
Obviously, and in contrast to the Heisenberg chain, $[H,j]
\neq 0$. Moreover, we find that the Hamiltonian's zero modes do not
contribute to the energy current. In general the current operator comprises
four eigenvalues, $j_1(k) = J_1 J_2 \sin(2k)/2$ and its own three zero
modes $j_{i \in \{2,3,4 \}}(k) = 0$. Interestingly, the non-zero modes
satisfy $H_i(k) \, \partial H_i(k)/ \partial k = - j_1(k)$.

\subsection{Energy-Current Autocorrelation}
\label{autocorrelation}

% ------------------------------- FIGURE 1 ----------------------------------
\begin{figure}[t]
\includegraphics[width=0.95\columnwidth]{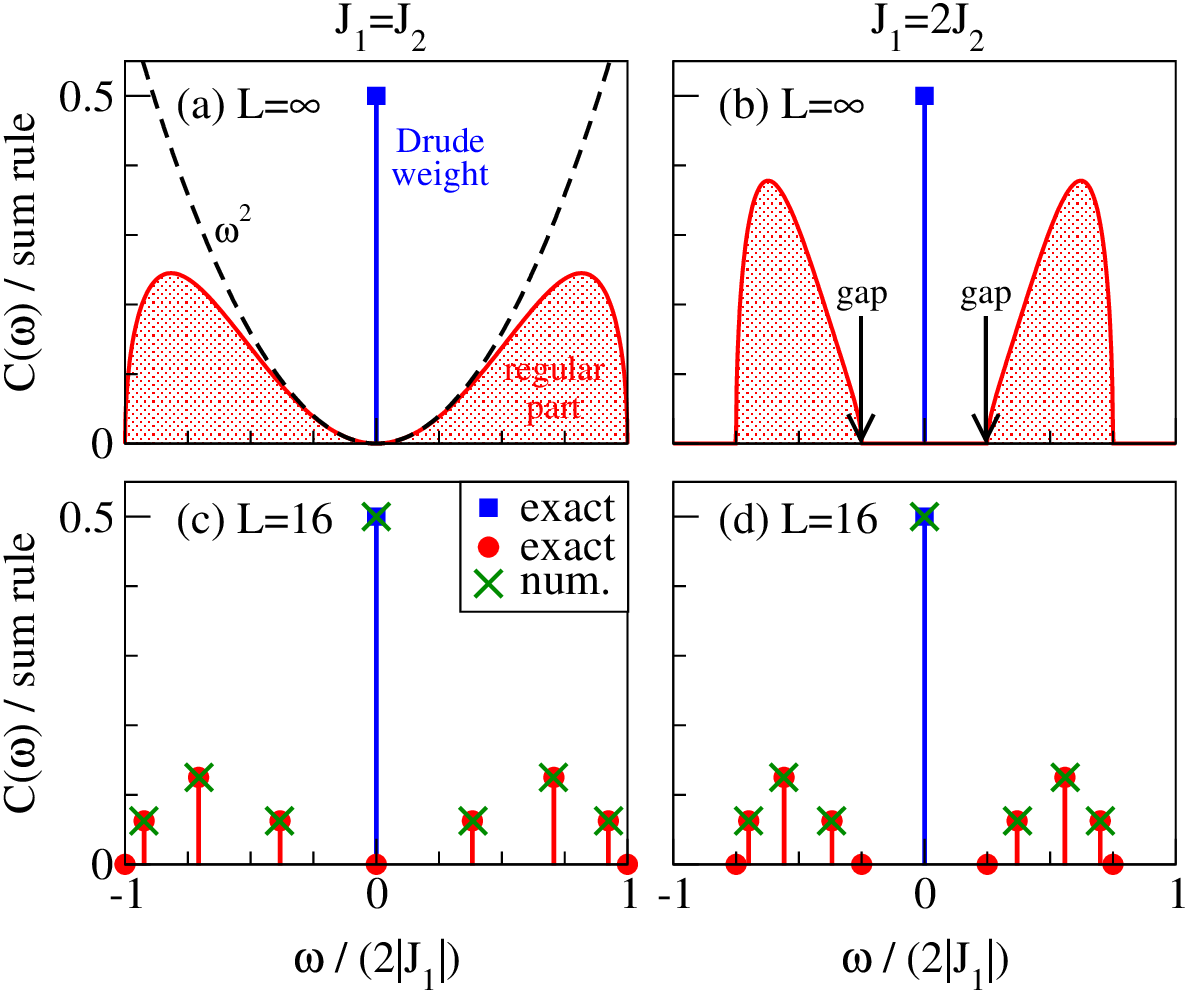}
\caption{(color online) Energy-current autocorrelation $C(\omega)$
for the Kitaev model at $\beta \to 0$ and (a) $J_1 = J_2$, (b) $J_1
= 2 J_2$ for $L=\infty$. (c), (d) show (a), (b) for $L=16$. In (b),
(d) the topological gap coexists with a nonzero Drude weight.}
\label{Fig1}
\end{figure}
% ---------------------------------------------------------------------------

In the following we investigate the energy-current autocorrelation
\begin{equation}
C(\omega) = \frac{1}{2 \pi \, L/2} \int_{-\infty}^\infty \text{d}t \,
\exp(-\imath \omega t) \, \langle j (t) j \rangle_\text{eq.}
\end{equation}
in frequency space $\omega$, where $\langle \ldots \rangle_\text{eq.}$
labels equilibrium averages at inverse temperatures $\beta = 1/T$ and
time arguments refer to the Heisenberg picture. The quadratic nature
of the theory allows for analytic results.

For simplicity, we discuss the case of $J_1 = J_2$ first and focus on
the high-temperature limit $\beta \rightarrow 0$. Here, $C(\omega)$ is a
symmetric function of $\omega$ and can be written at $\omega \neq 0$ as
\begin{equation}
C(\omega \neq 0) = \frac{1}{L/2} \sum_{k=0}^{\pi/2} A_k \sum_{\pm}
\delta(\omega \pm \omega_k)
\end{equation}
with the frequency $\omega_k = 2 |J_1| \cos k$ resulting from
the energy difference of the off-diagonal transitions $c_k^\dagger d_k$
and $d_k^\dagger c_k$. The amplitude $A_k = J_1^4 \sin(2k)^2/64$ is
essentially the square of the corresponding matrix elements. In the
thermodynamic limit $L \to \infty$, the $k$ sum can be converted into
an integral over the density of states $|\partial k/\partial \omega_k|$,
which can be carried out straightforwardly to yield the exact expression
\begin{equation}
C(\omega \neq 0) = \frac{|J_1|^3}{32 \pi} \sqrt{1-\Big(
\frac{\omega}{2J_1} \Big)^2} \Big( \frac{\omega}{2J_1} \Big)^2 \, .
\label{Eq4}
\end{equation}
Furthermore, since the integrated spectral weight $\int \text{d}\omega
\, C(\omega\neq 0) = J_1^4/128$ is exactly half of the $\beta \to 0$
`sum rule' $J_1^4/64$ (see Appendix \ref{sumrule}), we obtain
\begin{equation}
C(\omega) = C(\omega \neq 0) + \frac{J_1^4}{128} \, \delta(\omega) \, ,
\end{equation}
including a {\it nonzero} heat Drude weight.

Figure \ref{Fig1}~(a) summarizes the frequency-dependence of $C(\omega)$.
The nonzero Drude weight at zero frequency is commonly expected for
integrable models while counter examples exist (such as the vanishing spin
Drude weight in the gapped XXZ chain \cite{Zotos1999, FHM2003, Benz2005,
Prosen2011}). The existence and especially the form of the regular
part is a non-trivial property of the Kitaev model: While the
suppression of the regular part at low frequencies and in particular
the $\omega^2$-behavior has been proposed for all gapless models with
Drude weights \cite{Herbrych2012}, a rigorous proof is lacking and arguments
are based on numerical simulations of finite systems.

In the general case of $J_1 \neq J_2$, i.e., for a non-zero topological
gap, $C(\omega)$ can be obtained analogously. In Fig.~\ref{Fig1}~(b) we
show $C(\omega)$ at $J_1 = 2 J_2$. Clearly, the overall structure is similar
to Fig.~\ref{Fig1}~(a), but there is no spectral weight within a window of
$\delta \omega = J_1/2$, which resembles the topological gap between the
two branches of the one-particle dispersion. This gap coexists with a
nonzero Drude weight and vanishes at $J_1 = J_2$.

\section{Heisenberg-Kitaev model}
\label{heisenbergkitaev}

% ------------------------------- FIGURE 2 ----------------------------------
\begin{figure}[t]
\includegraphics[width=0.95\columnwidth]{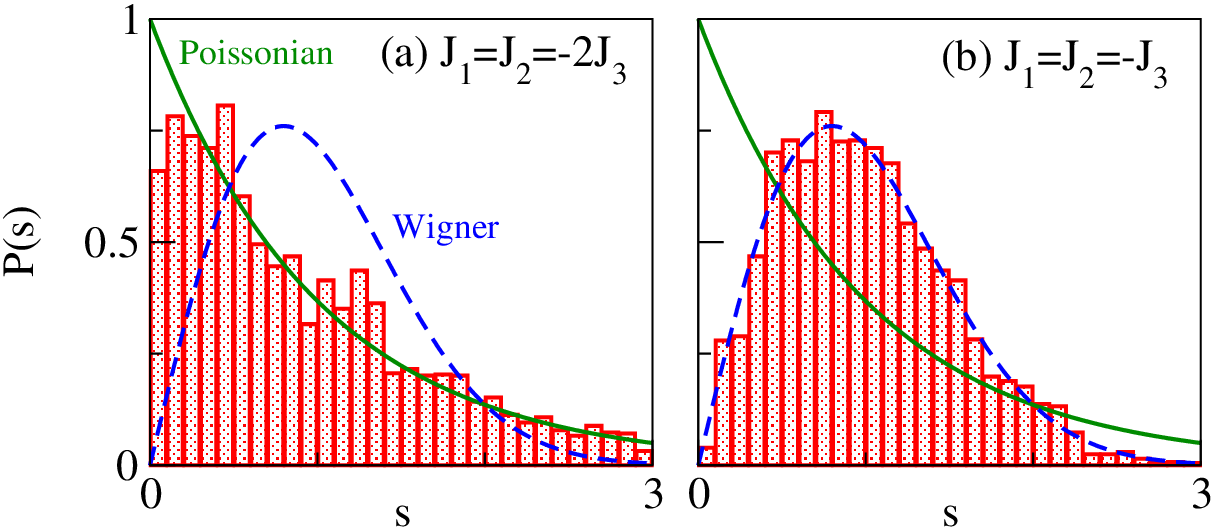}
\caption{(color online) Level-spacing distribution $P(s)$ for the
Heisenberg-Kitaev model at (a) $J_{i \neq 3} = -2J_3$, (b)
$J_{i\neq3} = -J_3$ for $L=16$ in a single symmetry sector. Curves
indicate both, the integrable Poissonian distribution and the quantum
chaotic Wigner distribution. Data for a wider range of model parameters
is shown in Appendix \ref{levelstatistics}.} \label{Fig2}
\end{figure}
% ---------------------------------------------------------------------------

% ------------------------------- FIGURE 3 ----------------------------------
\begin{figure}[b]
\includegraphics[width=0.95\columnwidth]{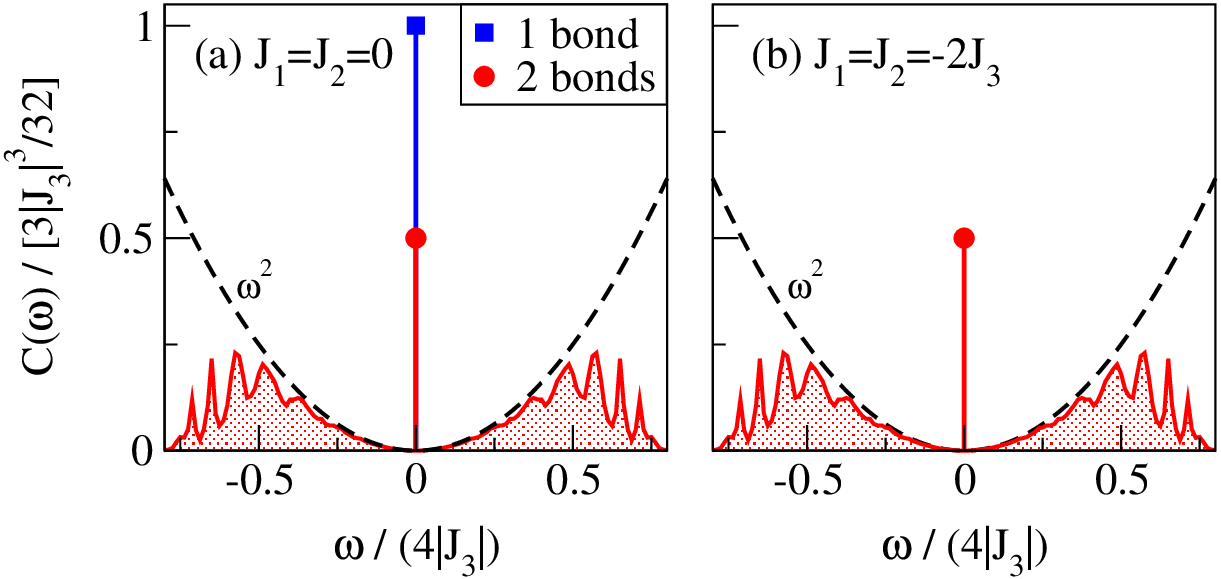}
\caption{(color online) Energy-current autocorrelation $C(\omega)$
at $\beta \to 0$ and (a) $J_{i\neq 3}=0$, (b) $J_{i\neq 3} =-2 J_3$
for $L=16$. At $\omega \neq 0$, $C(\omega)$ is averaged over
$\delta \omega = 0.05 \, |J_3|$.} \label{Fig3}
\end{figure}
% ---------------------------------------------------------------------------

% ------------------------------- FIGURE 4 ----------------------------------
\begin{figure}[t]
\includegraphics[width=0.95\columnwidth]{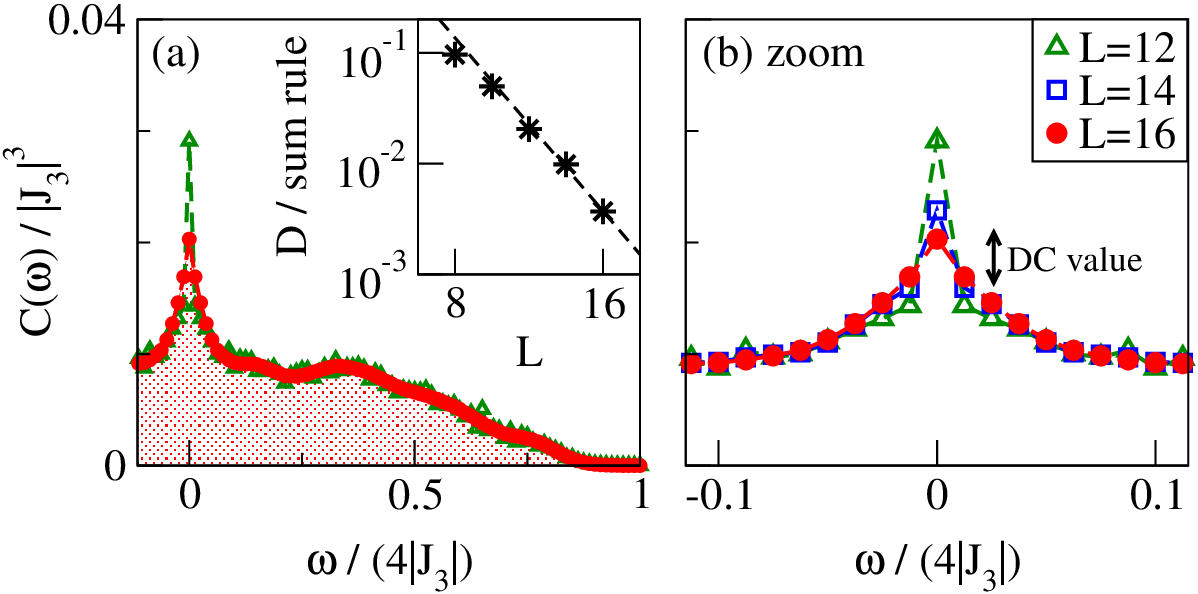}
\caption{(color online) Energy-current autocorrelation $C(\omega)$
for the Heisenberg-Kitaev model at $\beta \to 0$ and $J_{i\neq3} =
-J_3$ for $L=12$, $14$, and $16$. $C(\omega)$ is averaged over
$\delta \omega = 0.05 \, |J_3|$. (b) enlarges (a) at low frequencies.
The inset shows the finite-size scaling of the Drude weight in a
semi-log plot.} \label{Fig4}
\end{figure}
% ---------------------------------------------------------------------------

% ------------------------------- FIGURE 5 -----------------------------------
\begin{figure}[b]
\includegraphics[width=0.85\columnwidth]{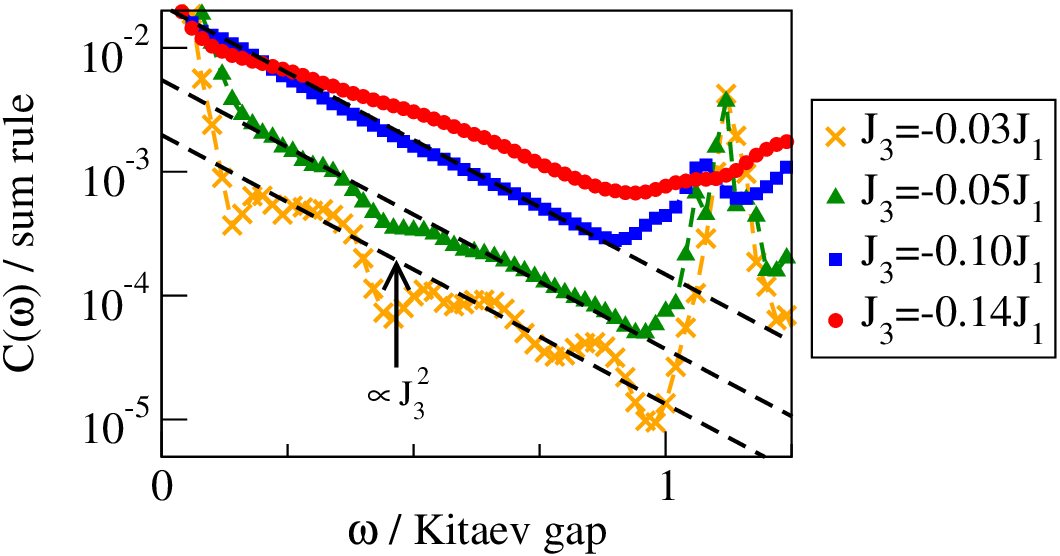}
\caption{(color online) Energy-current autocorrelation $C(\omega)$
for the Heisenberg-Kitaev model at $\beta \to 0$, $J_1 = 2 J_2$, and
small $J_3$ for $L = 16$ in a semi-log plot.} \label{Fig5}
\end{figure}
% ----------------------------------------------------------------------------

Now we add an XXZ type of Heisenberg exchange to Kitaev's model
\begin{equation}
H' = J_3 \sum_{r=1}^L S_r^x S_{r+1}^x + S_r^y S_{r+1}^y + \Delta \,
S_r^z S_{r+1}^z \, .
\end{equation}
This modification requires analytical approaches to be replaced
by numerical methods. We will use exact diagonalization of finite
systems and will focus on the antiferromagnetic and
isotropic case, $J_3 > 0$ and $\Delta = 1$, with a ferromagnetic
choice of $J_{i\neq3} < 0$.

While Kitaev's model is integrable and in 1D Heisenberg's model
is also in terms of the Bethe Ansatz, their sum $H + H'$ will
not be so in general. This is corroborated by the level-spacing
distribution $P(s)$, which is depicted in Fig.~\ref{Fig2}~(b).
It coincides with the Wigner distribution and thus
unveils a clear signature of quantum chaos, especially if $|
J_{i\neq3} | \sim |J_3|$. Note that a proper evaluation of
$P(s)$ requires an `unfolding' of the spectrum \cite{Reichl2004}
and a restriction to a single subspace of all `trivial' symmetries,
i.e., translation invariance, conservation of the even/odd particle
number, as well as particle-hole symmetry. (These symmetries do not
allow to reach the $L \sim 20$ sites for the Heisenberg chain.) Quantum
chaos prevails for all other ratios of $J_1$, $J_2$, $J_3$ we have
investigated, except for one additional integrable point at
$J_{i \neq 3} = -2 J_3$, see Fig.~\ref{Fig2}~(a). At this point,
local rotation in spin space \cite{Chaloupka2010} maps the
Heisenberg-Kitaev chain onto the Heisenberg chain only.

For the total Hamiltonian the energy current
\begin{eqnarray}
j &=& \sum_{l=1}^{L/2} \Big [(J_3 \! + \! J_1)(J_3 \! + \! J_2) \,
S_{2l}^yS_{2l+2}^x \! - \! J_3^2 \, S_{2l}^x S_{2l+2}^y \Big ]
S_{2l+1}^z \nonumber \\
&+& J_3 \, \Delta \Big [ J_3 \, S_{2l+1}^x S_{2l+2}^y \! - \! (J_3
\! + \! J_1) \, S_{2l+1}^y S_{2l+2}^x \Big ] S_{2l}^z \nonumber
\\[0.2cm]
&+& J_3 \, \Delta \Big [ J_3 \, S_{2l}^x S_{2l+1}^y \! - \! (J_3 \!
+ \! J_2) \, S_{2l}^y S_{2l+1}^x \Big ] S_{2l+2}^z
\label{Eq6}
\end{eqnarray}
turns into an operator with a rather involved structure, the
numerical implementation of which is delicate but can be checked by the
$\beta \to 0$ `sum rule' (see Appendix \ref{sumrule}). To further
check this, Fig.~\ref{Fig1}~(c), (d) shows that our exact diagonalization
perfectly reproduces the analytical result for $C(\omega)$ of the Kitaev
model. To this end, we have to compare numerically accessible chain
lengths with finite-size analytic results, featuring only
a few $\delta$ functions, rather than Eq.~(\ref{Eq4}). In the Kitaev limit
the finite-size spectrum is particularly sparse because the $L/2$ $Z_2$
invariants imply a $2^{L/2}$ fold degeneracy \cite{sen2010}.

Another important and non-trivial consistency check is $C(\omega)$ for the
Heisenberg model. In contrast to the conventional definition
\cite{Zotos1997}, the energy-current operator in Eq.~(\ref{Eq6}) is {\it not}
conserved, due to the two-site unit cell. While this emphasizes the
well-known ambiguity of defining local energy densities, it has {\it no}
consequence for the existence of a nonzero Drude weight at zero
frequency, as depicted in Fig.~\ref{Fig3}~(a). There, half of the total
spectral weight is concentrated in the Drude weight, similar to the Kitaev
model. Moreover, the regular part shows a striking similarity to the one
in Fig.~\ref{Fig1}~(a). As a final check, $C(\omega)$ at the
Heisenberg-equivalent additional integrable point at $J_{i\neq3}
=-2J_3$ turns out to be identical to that of the Heisenberg model, see
Fig.~\ref{Fig3}~(a) and (b).

Next we investigate $C(\omega)$ for the Heisenberg-Kitaev model within
the nonintegrable region, and in particular
for the case $J_{i\neq3} = -J_3$. In Fig.~\ref{Fig4} we show
$C(\omega)$ at $\beta \rightarrow 0$ for finite chains of length $L=12$,
$14$, and $16$. Several comments are in order. First, at $\omega \gg 0$,
$C(\omega)$ is independent of $L$ and a smooth function of $\omega$, at least
at a reasonable scale $\delta \omega = 0.05 \, |J_3|$. Second, at $\omega = 0$,
$C(\omega)$ still depends on $L$ due to the presence of a nonzero Drude
weight in finite chains; however, as shown in the inset of Fig.~\ref{Fig4},
this Drude weight decreases exponentially with $L$. Such finite-size
scaling is commonly expected for nonintegrable quantum many-body systems
\cite{steinigeweg2013}. Third, although the overall structure of $C(\omega)$
is broad, a narrow `peak' is clearly visible around zero frequency.
Certainly, this peak may be interpreted as the broadening of the $\delta$
function, which is rather separated from the regular part at the three
integrable points of the Heisenberg-Kitaev model, cf.~Fig.~\ref{Fig1}~(a)
and Fig.~\ref{Fig3}~(a), (b). The width of the `peak' turns out to increase
with $\Delta$, although not shown explicitly here.

The situation is similar for $J_1 \neq J_2$. In particular, there is no
low-frequency suppression of the regular part at intermediate $J_3$. Yet,
at small $J_3$, we can clearly identify the topological gap even at high
temperatures, see Fig.~\ref{Fig5}. The gap is only weakly dependent on
$J_3$ up to $J_3 = -0.14 J_1$, beyond which all signatures of the gap are
hidden by high-temperature excitations $\propto J_3^2$, as expected from
perturbation theory.

% ------------------------------- FIGURE 6 -----------------------------------
\begin{figure}[t]
\includegraphics[width=0.75\columnwidth]{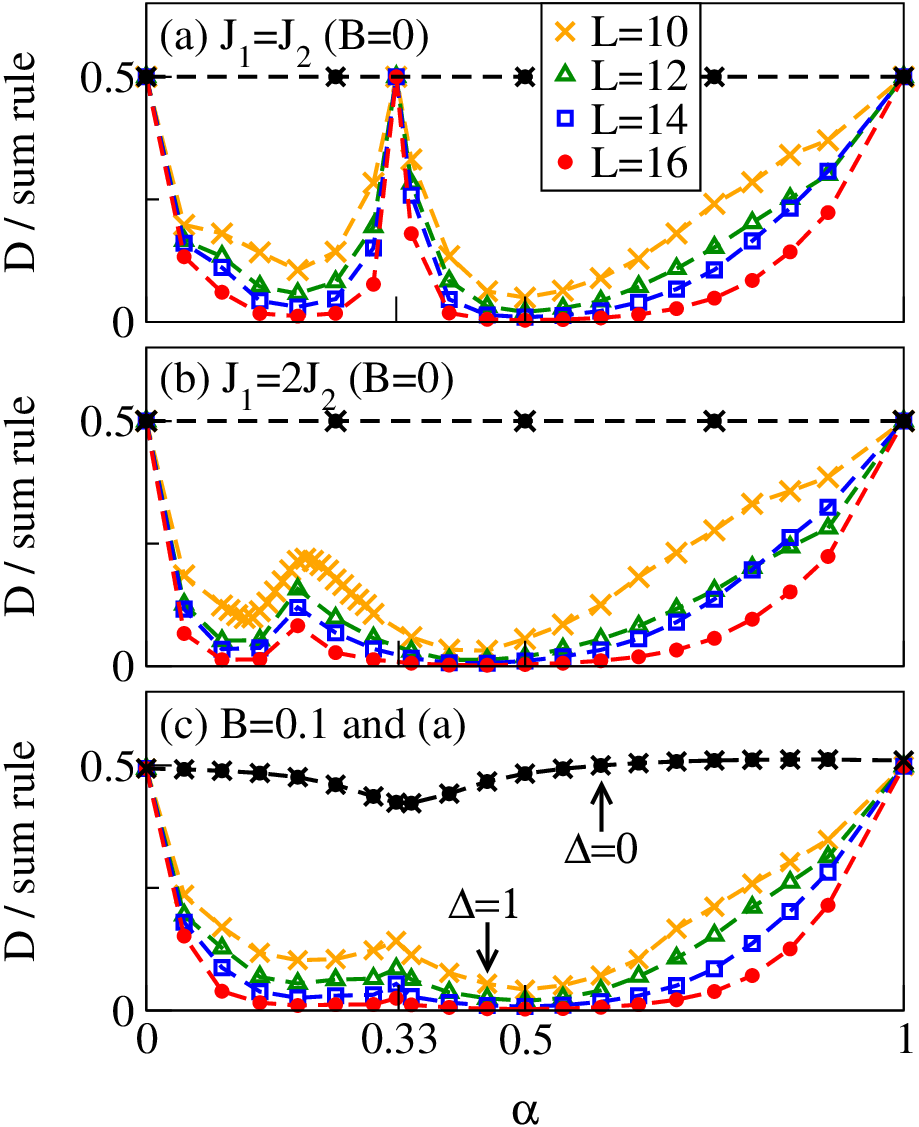}
\caption{(color online) High-temperature Drude weight of the
Heisenberg-Kitaev model at $J_3 = \alpha$, $J_1 = \alpha-1$, and
(a) $J_1 = J_2$, (b) $J_1 = 2 J_2$ for $L=10$, $12$, $14$, and
$16$. (c) is identical to (a) with $B \neq 0$.}
\label{Fig6}
\end{figure}
% ----------------------------------------------------------------------------

We now discuss the $J_1$-$J_2$-$J_3$ dependence of the Drude weight
in detail. To this end, we introduce a parameter $\alpha \in [0,1]$, rewriting
the exchange coupling constants as $J_3 = \alpha$ and $J_{i\neq3}=
\alpha-1$. In turn, the Kitaev model is realized for $\alpha=0$ and the Heisenberg
model for $\alpha = 1$. Figure \ref{Fig6}~(a) shows the high-temperature Drude
weight vs.~$\alpha$ for finite chains of length $L=10$, $12$, $14$, and $16$.
Clearly, the Drude weight takes on its minimum value at $\alpha \sim 1/2$. This
minimum is merely a small fraction of the `sum rule' for $L=16$ and, in
view of the inset of Fig.~\ref{Fig4}, it clearly approaches zero in the
thermodynamic limit $L \to \infty$. In the immediate vicinity of the
integrable points at $\alpha = 0$, $1/3$, or $1$ such conclusions are less
obvious. While the numerical data may be consistent with nonzero
Drude weights $D > 0$ in the thermodynamic limit $L \to \infty$, the system
sizes are too small and would also be consistent with $D=0$. Note that for $J_1
\neq J_2$ the peak at the point $\alpha = 1/3$ is shifted to other values of
$\alpha$ and, as a consequence of nonintegrability, vanishes for $L \to
\infty$, see Fig.~\ref{Fig6}~(b).

As an interesting side remark, for extreme anisotropy $\Delta = 0$,
the parameter $\alpha$ tunes from free Majorana to free XY fermions.
In that case, as shown in Fig.~\ref{Fig6}~(a), (b), the Drude weight is
independent of $\alpha$, corroborating a picture of free particles
for all $\alpha$. We note that results on $0 < \Delta < 1$ are
similar to the $\Delta = 1$ results and shown in Appendix
\ref{smalldelta}.

\section{External Magnetic Field}
\label{magneticfield}

Finally, we also consider a $z$ axis oriented external magnetic
field $B$ by including a Zeeman energy $H'' = B S^z$. This leads to a
magneto-thermal contribution $j''$ to the energy current
\begin{eqnarray}
j'' &=& \sum_{l=1}^{L/2} \frac{B}{2} \Big [ J_3 \, S_{2l+1}^x
S_{2l+2}^y \! - \! (J_3 \! + \! J_1) \, S_{2l+1}^y S_{2l+2}^x
\Big ] \nonumber \\
&+& \frac{B}{2} \Big [ J_3 \, S_{2l}^x S_{2l+1}^y \! - \! (J_3 \! +
\! J_2) \, S_{2l}^y S_{2l+1}^x \Big ] \, ,
\end{eqnarray}
which is a two-site operator and simplifies to the well-known
{\it spin} current at $J_{i\neq3} = 0$ \cite{Zotos1999, FHM2003,
Benz2005, Prosen2011}.

We choose a `small' $B=0.1$ and repeat the calculation of the Drude weight
in Fig.~\ref{Fig6}~(a) for $J_3 = \alpha$ and $J_{i\neq3} = 1-
\alpha$. The result of the calculation is depicted in Fig.~\ref{Fig6}~(c).
Apparently, the Drude weight at $\alpha=1/3$ is extremely sensitive to
$B$ and seems to vanish in the thermodynamic limit $L \to \infty$. In
sharp contrast, the Drude weight is much less sensitive for free particles
at $\Delta = 0$ and remains nonzero for $L \to \infty$.

\section{Conclusion}
\label{conclusion}

In summary, we have investigated the energy dynamics in the Heisenberg-Kitaev
chain, representing a model which interpolates between two qualitatively different
integrable points in one dimension. To this end, we have used analytical
and numerical methods within linear response theory. Focusing 
on the limit of high temperatures, we have varied several parameters of the
model including the strength, ratio, and anisotropy of exchange coupling as
well as the external magnetic field.

As a central result, we have shown that all integrable points display nonzero
Drude weights at frequency $\omega = 0$ and suppressed regular contributions
at low frequencies $\omega \neq 0$, with a quadratic frequency dependence 
$\propto \omega^2$ for the gapless cases considered. For the bare Kitaev
chain, our analytical result for the current-current correlation in the
thermodynamic limit has proven the coexistence of a nonzero Drude weight
and the topological gap. Our numerical analysis has further unveiled signatures
of the topological gap in the response function of Kitaev chains perturbed
by XXZ exchange. Remarkably, we have found numerical evidence that these
signatures can still be seen for rather strong perturbations. Off integrability,
our numerical results have indicated extended quantum chaotic regions with
vanishing Drude weights at frequency $\omega = 0$ and well-defined conductivities
in the DC limit $\omega \to 0$.

Although we have focused on high temperatures and one dimension, we
believe that our results could be a first relevant step towards theoretical
understanding of transport in the Heisenberg-Kitaev model on 2D lattices
and an understanding of transport experiments in novel local-moment materials
with strong spin-orbit coupling. Future directions of research thus include
lower temperatures and higher dimensionality, with quasi-1D ladder structures
as an obvious next step. In this context it is important to note that lower
temperatures typically go along with significantly larger finite-size effects
and that higher dimensionality reduces the edge length of the lattice accessible
by exact diagonalization. It is therefore indispensable to increase system size
beyond the $16$ sites studied in our work and to replace exact diagonalization
by other numerical techniques. Such a technique may be dynamical quantum
typicality, which has been applied recently to study Drude weights
\cite{steinigeweg2014} and dc conductivities \cite{steinigeweg2016} of
low-dimensional spin systems.

% ------------------------------- FIGURE 7 ----------------------------------
\begin{figure}[t]
\includegraphics[width=0.95\columnwidth]{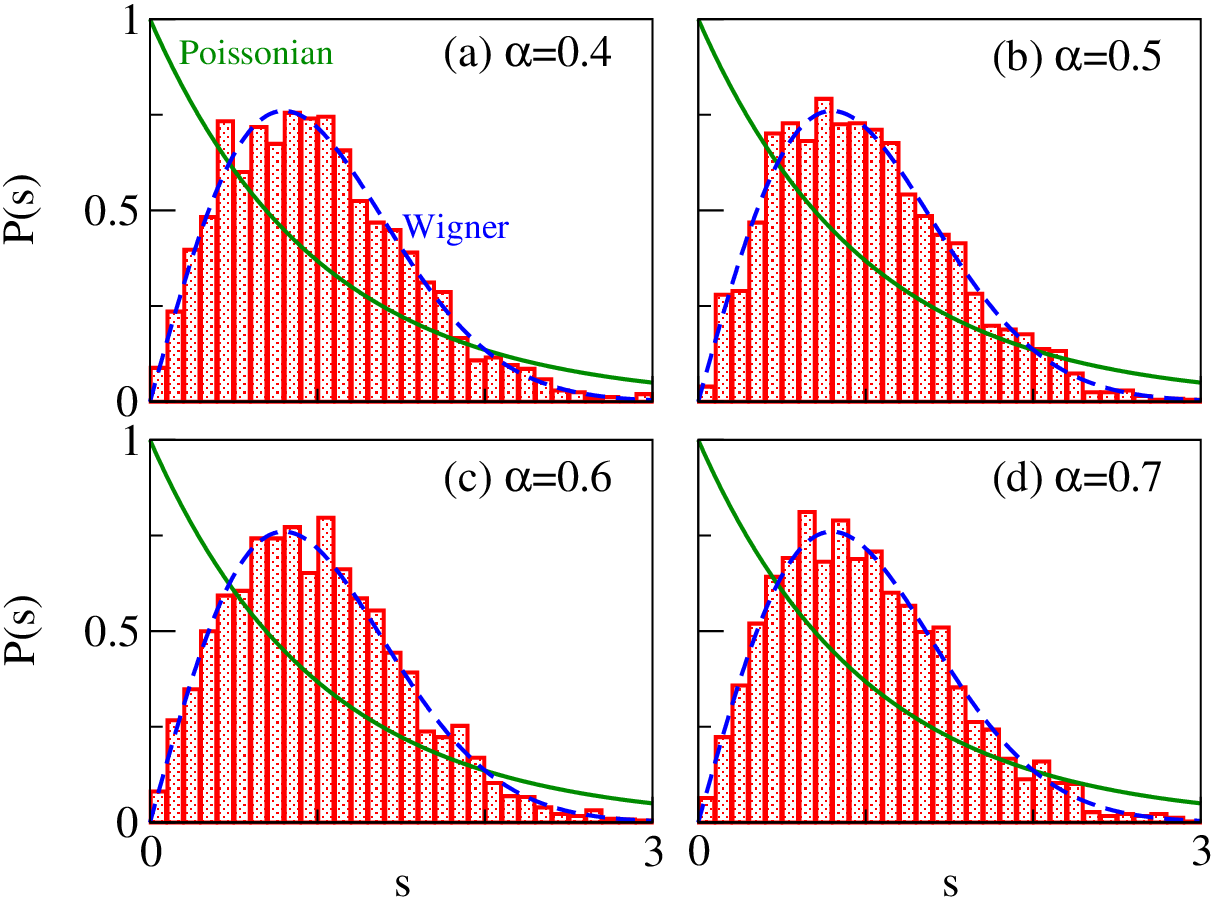}
\caption{(color online) Level-spacing distribution $P(s)$ for the
Heisenberg-Kitaev chain at four different exchange-coupling constants $J_3
= \alpha$, $J_1 = J_2 = 1 - \alpha$: (a) $\alpha = 0.4$, (b) $\alpha = 0.5$,
(c) $\alpha = 0.6$, and (d) $\alpha = 0.7$. While bars represent numerical
data for $L=16$ and a single symmetry sector, curves indicate both, the
integrable Poissonian distribution and the quantum chaotic Wigner distribution.
Clearly, quantum chaos emerges at all $\alpha$ depicted.}
\label{Fig7}
\end{figure}
% ---------------------------------------------------------------------------

% ------------------------------- FIGURE 8 -----------------------------------
\begin{figure}[t]
\includegraphics[width=0.75\columnwidth]{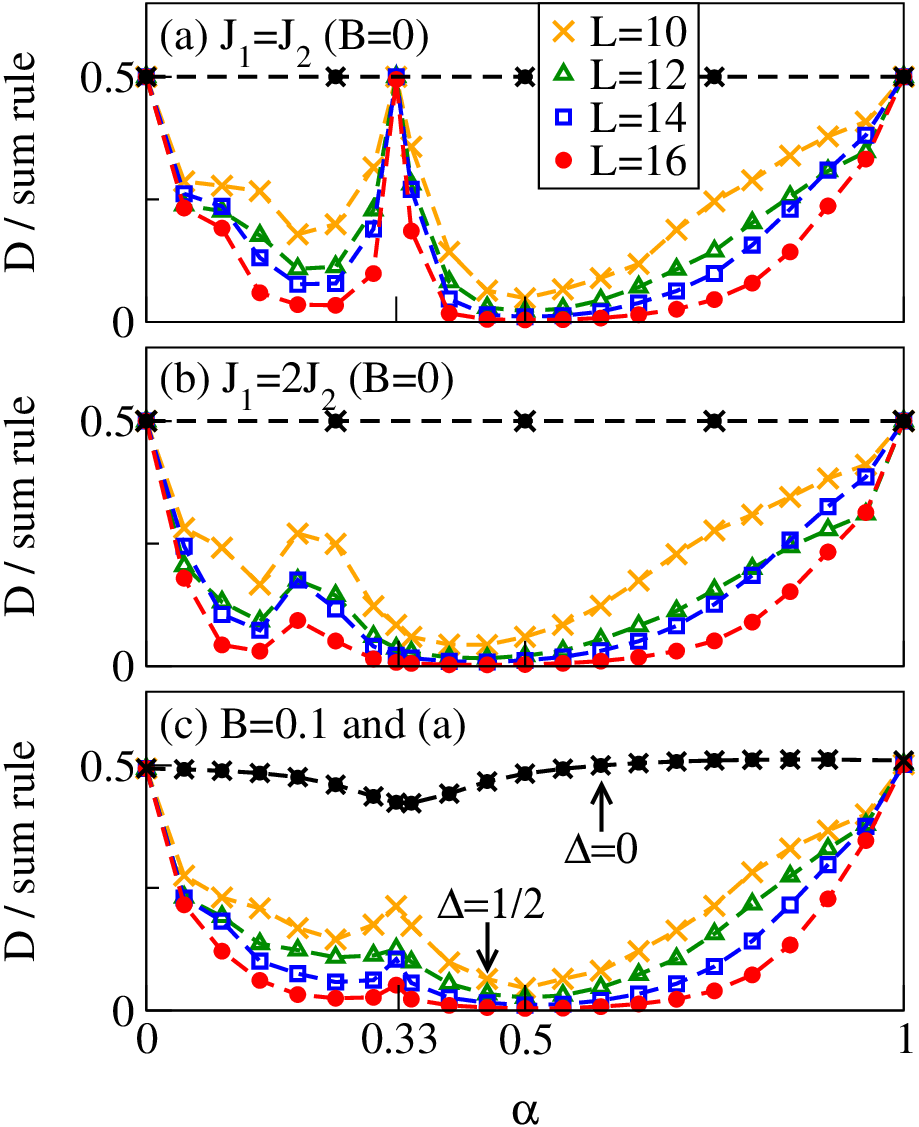}
\caption{(color online) High-temperature Drude weight of the $\Delta < 1$
Heisenberg-Kitaev model at $J_3 = \alpha$, $J_1 = \alpha-1$, and (a) $J_1
= J_2$, (b) $J_1 = 2 J_2$ for $L=10$, $12$, $14$, and $16$. (c) is identical
to (a) with $B \neq 0$. Data for $\Delta =1/2$ is similar to $\Delta = 1$
data in the paper.}
\label{Fig8}
\end{figure}
% ----------------------------------------------------------------------------

\acknowledgements

We sincerely thank P.~Recher and L.~Weithofer for fruitful discussions.
This work has been supported in part by DFG FOR912 Grant No.~BR 1084/6-2,
EU MC-ITN LOTHERM Grant No.~PITN-GA-2009-238475, and DFG SFB 1143. Wolfram
Brenig also acknowledges kind hospitality of the PSM Dresden.

\appendix

\section{Extended Quantum Chaotic Regions}
\label{levelstatistics}

In the paper we have numerically demonstrated that the nearest-neighbor
level-spacing distribution $P(s)$ for the Heisenberg-Kitaev chain at
exchange-coupling constants $J_1 = J_2 = -J_3$ agrees with the Wigner
distribution. In this way we have unveiled a clear signature of quantum
chaos and hence the nonintegrability of the model, at least at a specific
point in parameter space. To demonstrate that this point lies in an extended
quantum chaotic region, we present in Fig.\ \ref{Fig7}\ additional results
on $P(s)$ for several $J_3 = \alpha$, $J_1 = J_2 = 1 - \alpha$. Evidently,
quantum chaos prevails for all $\alpha$ depicted. As discussed in the paper,
at these $\alpha$ the energy Drude weight clearly vanishes in thermodynamic
limit.

\section{Drude Weights for $0 < \Delta < 1$}
\label{smalldelta}

In the paper we have shown that at extreme exchange anisotropy $\Delta
= 0$ the energy Drude weight is nonzero in the thermodynamic limit, for all
exchange-coupling constants $J_1$, $J_2$, and $J_3$ as well as for zero and
nonzero external magnetic field $B$. Contrary, our numerical results have
indicated that at $\Delta = 1$ the energy Drude weight is only nonzero at
the three integrable points at $\alpha = 0$, $1/3$, and $1$ ($J_1 = J_2$ and
$B \neq 0$ for $\alpha = 1/3$). To demonstrate that our $\Delta = 1$ findings
are also representative for smaller values of $\Delta$, we repeat in Fig.\
\ref{Fig8} our calculations for the intermediate value $\Delta = 1/2$.
Apparently, data for $\Delta = 1/2$ is similar to $\Delta = 1$ data in the
paper.

\section{Sum Rule}
\label{sumrule}

The $\beta \to 0$ `sum rule' for the most general form of $j$ reads
\begin{eqnarray}
\frac{\text{tr}\{ j^2 \}}{2^L \, L} &=& \frac{(J_3^2 \Delta^2 + B^2)
\, [2J_3^2 + (J_3 \! + \! J_1)^2 + (J_3 \! + \! J_2)^2]} {128} \nonumber \\
&+& \frac{J_3^4 + (J_3 \! + \! J_1)^2 (J_3 \! + \! J_2)^2}{128}
\label{trace}
\end{eqnarray}
and is convenient to verify the numerical implementation of this operator. While
$\beta \gg 0$ is not studied in our paper, it is worth mentioning that the sum
rule at such $\beta$ differs from Eq.\ (\ref{trace}).

\newpage

\end{document}